\documentclass[conference]{IEEEtran} 
\usepackage{amsmath}
\usepackage{amsmath,amssymb,amsfonts}
\usepackage{cite}
\usepackage{mathtools}
\usepackage[binary-units=true]{siunitx}
\DeclareSIUnit{\belmilliwatt}{Bm}
\usepackage{amsfonts}
\usepackage{amssymb}
\usepackage{graphicx}
\usepackage{dsfont}
\usepackage{subfigure}
\usepackage{algorithm}
\usepackage{algpseudocode}
\usepackage{mathtools}

\usepackage{textcomp}
\usepackage[export]{adjustbox}
\usepackage{float}
\usepackage{booktabs}
\usepackage{multicol}
\usepackage{xparse}
\usepackage{color,soul}
\usepackage[bottom]{footmisc}
\usepackage[binary-units=true]{siunitx}
\DeclareSIUnit{\belmilliwatt}{Bm}
\DeclareSIUnit{\dBm}{\deci\belmilliwatt}
\DeclareSIUnit[per-mode=symbol,per-symbol=p]{\Bps}{\byte\per\second}

\sethlcolor{yellow}
\usepackage{algpseudocode}

\usepackage{hyperref}

\makeatletter
\def\BState{\State\hskip-\ALG@thistlm}
\makeatother

\begin{document}

    \title{On the Analysis of AoI-Reliability Tradeoff in Heterogeneous IIoT Networks 
}

\author{
\IEEEauthorblockN{Hossam Farag, Syed Muhammad Ali and \v{C}edomir Stefanovi\'{c}}
\IEEEauthorblockA{
Department of Electronic Systems, Aalborg University, Denmark\\
Email: \{hmf, syedma, cs\}@es.aau.dk}}

	\maketitle
	\begin{abstract}
Age of information (AoI) and reliability are two critical metrics to support real-time applications in Industrial Internet of Things (IIoT). These metrics reflect different concepts of timely delivery of sensor information. Monitoring traffic serves to maintain fresh status updates, expressed in a low AoI, which is important for proper control and actuation actions. On the other hand, safety-critical information, e.g., emergency alarms, is generated sporadically and must be delivered with high reliability within a predefined deadline. In this work, we investigate the AoI-reliability trade-off in a real-time monitoring scenario that supports two traffic flows, namely AoI-oriented traffic and deadline-oriented traffic. Both traffic flows are transmitted to a central controller over an unreliable shared channel. We derive expressions of the average AoI for the AoI-oriented traffic and reliability, represented by Packet Loss Probability (PLP), for the deadline-oriented traffic using Discrete-Time Markov Chain (DTMC). We also conduct discrete-event simulations in MATLAB to validate the analytical results and evaluate the interaction between the two types of traffic flows. The results clearly demonstrate the tradeoff between the AoI and PLP in such heterogeneous IIoT networks and give insights on how to configure the network to achieve a target pair of AoI and PLP.  
	\end{abstract}
\section{Introduction}\label{sec:intro}

Industrial Internet of Things (IIoT) is a key pillar for the Industry 4.0 paradigm to enable smart manufacturing \cite{IIoT}.
In IIoT applications, the field network, realized by Industrial Wireless Sensor Network (IWSN), allows remote monitoring of wide-ranging industrial processes, thereby enabling efficient and sustainable production. 
IWSNs typically serve the communication of different traffic flows that are characterized  by different generation patterns (event-triggered and time-triggered) and communication requirements in terms of latency, throughput and timeliness~\cite{HeT-traffic}. 
In process monitoring and control scenarios, emergency alarms and safety-critical information represent the event-triggered flows that must be transmitted within stringent deadline constraints to maintain system stability and avoid dangerous consequences. 
On the other hand, regular, periodic monitoring traffic represents the time-triggered traffic of sensor readings to be transmitted to the central controller.
The freshness of such traffic is crucial to keep the central controller updated with the status of the industrial process to drive decisions or feedback loops. The information freshness however, cannot be captured by traditional performance metrics, such as delay and throughput. Age of Information (AoI)~\cite{AoI} was introduced as a relevant metric for quantifying information freshness from the perspective of the central controller. AoI is defined as the time elapsed since the latest received packet was generated, thereby it is intrinsically different from classical performance metrics like throughput and delay, which focuses on a single packet and only captures its network time. It has been proven that the adopted strategies to minimize delay or to maximize throughput are not necessarily optimal for minimizing AoI~\cite{delay-throughput}. 

Several research works were devoted to the study and improvement of the performance of IWSNs with heterogeneous traffic \cite{HeT-traffic, HeT1, HeT2, HeT3}.
The main goal of these approaches is to give critical traffic the highest transmission priority while sacrificing the delivery and reliability of less-critical traffic.
However, none of these works consider the AoI performance, where the freshness of less critical traffic might be subject to AoI constraints.
In \cite{AoI-sensor1}, the authors propose improved multiple access schemes to improve the AoI performance of energy harvesting IWSNs.
The authors in \cite{AoI-sensor2} study the AoI performance in UAV-aided IWSNs where they formulate two optimization problems to minimize the average and maximal AoI.
The work in \cite{AoI-sensor3} presents two greedy scheduling policies to minimize the AoI and jitter in industrial cyber-physical systems. All these works focus on analyzing or optimizing the performance of AoI in IWSNs where the network supports only a single traffic flow of AoI-oriented data, which is not the case in most IIoT applications. In heterogeneous IIoT networks, different traffic flows are associated with different performance metrics and their corresponding sources contend over unreliable multi-access channel.
The authors in \cite{AoI-HeT} propose an optimal generation policy for the  status updates in a heterogeneous IoT network that serves AoI-sensitive and AoI-insensitive traffic flows, however, this work focuses only on the optimal AoI performance of the AoI-sensitive traffic and disregards the performance of the AoI-insensitive one. In typical IIoT applications, the AoI-insensitive flow could be a critical traffic that is characterized by stringent deadline constraint, and its reliability (i.e., the ratio of packets delivered within the deadline bound) is crucial for the stability and functionality of the system.    

In this work, we investigate the AoI-reliability tradeoff in heterogeneous IIoT network that supports two traffic flows, namely AoI-oriented traffic and deadline-oriented traffic. The packets from both traffic flows are transmitted to a common central controller via unreliable multi-access channel. We derive the average AoI for the AoI-oriented traffic and the reliability, represented by Packet Loss Probability (PLP), for the deadline-oriented traffic using Discrete-Time Markov Chain (DTMC). We also validate our analysis through discrete-event simulations via MATLAB. The obtained results demonstrate the interaction between the two traffic flows and show that the overall network performance  is mainly influenced by the access probabilities. The analysis and the results in this work give insights on  how to configure the system to achieve a target pair of AoI and PLP, which could be through adopting enhanced channel access and/or queue management strategies. 

The remainder of the text is organized as follows. Section~\ref{network-model} describes the network model and the basic system parameters. Section~\ref{sec:AoI-analysis} presents the DTMC analysis of the AoI, followed by the DTMC analysis of the queue size and PLP in Section~\ref{sec:PLP-analysis}. The results are given in Section~\ref{sec:results}, and finally the paper is concluded in Section~\ref{sec:conclusions}. 

\section{Network Model}
\label{network-model}

We consider a static deployment of an IWSN in a process monitoring scenario where a set of sensor nodes are randomly distributed to monitor an industrial process.
All sensory information is transmitted to a central controller using a time-slotted random access via a fading channel. 
Specifically, we consider a slow and flat Rayleigh fading channel with additive white Gaussian noise within a time slot.
The network supports two types of traffic flows, namely deadline-oriented traffic~($\mathrm{T_{D}}$) and AoI-oriented traffic~($\mathrm{T_{AoI}}$).
The $\mathrm{T_{D}}$ traffic represents alarms and safety-critical data that are generated at emergency events and must be delivered to the central controller within a predefined deadline.
The $\mathrm{T_{AoI}}$ traffic represents time-triggered sensory data where the goal is to keep the central controller’s received information as fresh as possible. 
Without loss of generality, we consider one node that transmits $\mathrm{T_{D}}$ traffic and $N$ nodes transmitting the $\mathrm{T_{AoI}}$ traffic.
At each time slot, a packet arrives to the infinite buffer of the $\mathrm{T_{D}}$ with probability $\lambda$. Further, each $\mathrm{T_{D}}$ packet is attached with a constant deadline of $D$, such that it should be delivered within $D$ time slots since its arrival, otherwise it is dropped.
At the beginning of a time slot, the $\mathrm{T_{D}}$ node (if its queue is non-empty) attempts to transmit the packet at the head of the queue with probability $p_1$. When the transmission of the $\mathrm{T_{D}}$ fails, the packet is retransmitted until it is either successfully received or its deadline has expired.
The $\mathrm{T_{AoI}}$ traffic follows the generate-at-will model \cite{AoI-model}, in which a $\mathrm{T_{AoI}}$ node generates a fresh sample when it decides to transmit, i.e., access the channel with probability $p_2$. After the transmission attempt, the $\mathrm{T_{AoI}}$ node discards the packet, i.e., there is no retransmissions.
All such nodes are assumed to be synchronized, and the packet arrivals align with the boundary of the time slot. Acknowledgments of successful transmission are received within the same time slot via an error-free channel. 

We consider the capture effect~\cite{capture}, where the central controller can successfully decode a packet if the received Signal-to-Interference-plus-Noise Ratio (SINR) exceeds a certain threshold~$\gamma$ (capture ratio). The threshold~$\gamma$ is determined according to a certain packet error probability as a function of packet length, modulation, channel coding, diversity and receiver design~\cite{SINR-TH}.
Let $S$ denotes the set of nodes concurrently transmitting within the same time slot, the $\text{SINR}_{i}$ at the central controller corresponding to an arbitrary transmitting node $i$ is 
\begin{equation}\label{SINR}
\text{SINR}_{i}= \frac{P_i|h_{i,c}|^{2}d_{i,c}^{-\alpha}}{\sigma^2+\sum_{j\in S\setminus\{i\}}P_j|h_{j,c}|^{2}d_{j,c}^{-\alpha}},  
\end{equation}
where $P_i$ is the transmitting power of node $i$, $h_{i,c}$ is the Rayleigh random variable of the channel between node $i$ and the central controller $c$ ($|h_{i,c}|^{2}$ is exponentially distributed \cite{Rayleigh}), $d_{i,c}$ denotes the distance between the node $i$ and the central controller, $\sigma^2$ is the noise power and $\alpha$ is the path loss exponent.
In order for the central controller to successfully decode a received packet, it should be $\text{SINR}_i>\gamma$ and the probability of this event is given as 
\begin{equation}\label{PS}
\begin{split}
 \mathds{P}(\text{SINR}_i>\gamma)&=\mathrm{exp}\left(-\frac{\gamma \sigma^2}{P_i|h_{i,c}|^{2}d_{i,c}^{-\alpha}}\right)\times\\
 &\prod_{j\in S\setminus\{i\}}\left(1+\gamma \frac{P_j|h_{j,c}|^{2}d_{j,c}^{-\alpha}}{P_i|h_{i,c}|^{2}d_{i,c}^{-\alpha}}\right)^{-1}.
\end{split}    
\end{equation}
Based on \eqref{PS}, the successful update probability per time slot $q_{D}$ of the $T_D$ node, i.e, the probability of successfully receiving a $T_D$ packet within a time slot, is given as
\begin{equation}
q_{D}=\sum_{k=0}^{N}\binom{N}{k}p_1(1-p_2)^{N-k}{p_2}^k P_{S_k},
\end{equation}
where $P_{S_k} = \mathds{P}(\text{SINR}_i>\gamma\,\,|\,\, |S|=k)$, which can be obtained via \eqref{PS}. 

The successful update probability $q_{{AoI}}$ of a $\mathrm{T_{AoI}}$ node, which reflects a successful delivery of an update at the end of a time slot, depends on the queue status of the $T_D$ node. Let $Q$ be a random variable that represents the status of the output queue of the $T_D$ node (empty or non empty), then  $q_{{AoI}}$ can be expressed as   
\begin{equation}\label{service-AoI}
\begin{split}
 q_{{AoI}}&=p_2[\underbrace{q_{S_0}\mathds{P}(Q=0)}_{L_1}+\underbrace{q_{S_0}(1-p_1)\mathds{P}(Q>0)}_{L_2}\\
&+\underbrace{q_{S_1}p_1\mathds{P}(Q>0)}_{L_3}],  
\end{split}
\end{equation}
where $q_{S_0}$ and $q_{S_1}$ represent the successful probabilities when the $\mathrm{T_{D}}$ node is idle (not transmitting) and active (transmitting), respectively. The successful update probability $q_{{AoI}}$ in \eqref{service-AoI} is calculated by three terms corresponding to three different cases. The term $L_1$ denotes an empty queue of the $\mathrm{T_{D}}$ node, thus not transmitting in the current time slots, and only the $\mathrm{T_{AoI}}$ nodes attempting to transmit. The term $L_2$ refers to non-empty queue of the $\mathrm{T_{D}}$ node, but it decides not to transmit in the current time slot. Finally, the term $L_3$ represents the case when the $\mathrm{T_{D}}$ node decides to transmit the packet at the head of the queue and contend with the attempting $\mathrm{T_{AoI}}$ nodes.    
\begin{equation}
q_{S_0}=\sum_{k=0}^{N-1}\binom{N-1}{k}(1-p_2)^{N-k-1}{p_2}^k P_{S_k},    
\end{equation}
\begin{equation}
q_{S_1}=\sum_{k=0}^{N-1}\binom{N-1}{k}(1-p_2)^{N-k-1}{p_2}^k P_{S_{k+1}}.    
\end{equation}
\section{Analysis of the AoI} \label{sec:AoI-analysis}
In this section, we analyze the average AoI at the central controller corresponding to the $\mathrm{T_{AoI}}$ traffic.
We assume that the status updates from  all $\mathrm{T_{AoI}}$ nodes are equally important.
Therefore, without loss of generality, we consider an arbitrary $\mathrm{T_{AoI}}$ node and evaluate the AoI in discrete time.
The AoI represents the number of time slots elapsed since the last received packet was generated. If $\Delta (t)$ denotes the AoI at the end of time slot $t$, then we have  
\begin{equation}
\Delta (t+1)=\begin{cases}
\Delta (t)+1 & \text{unsuccessful transmission}\\
1 & \text{successful transmission}.
\end{cases}
\end{equation}
    \begin{figure}[t!] 
		\centering
		\includegraphics[width= 0.7\linewidth]{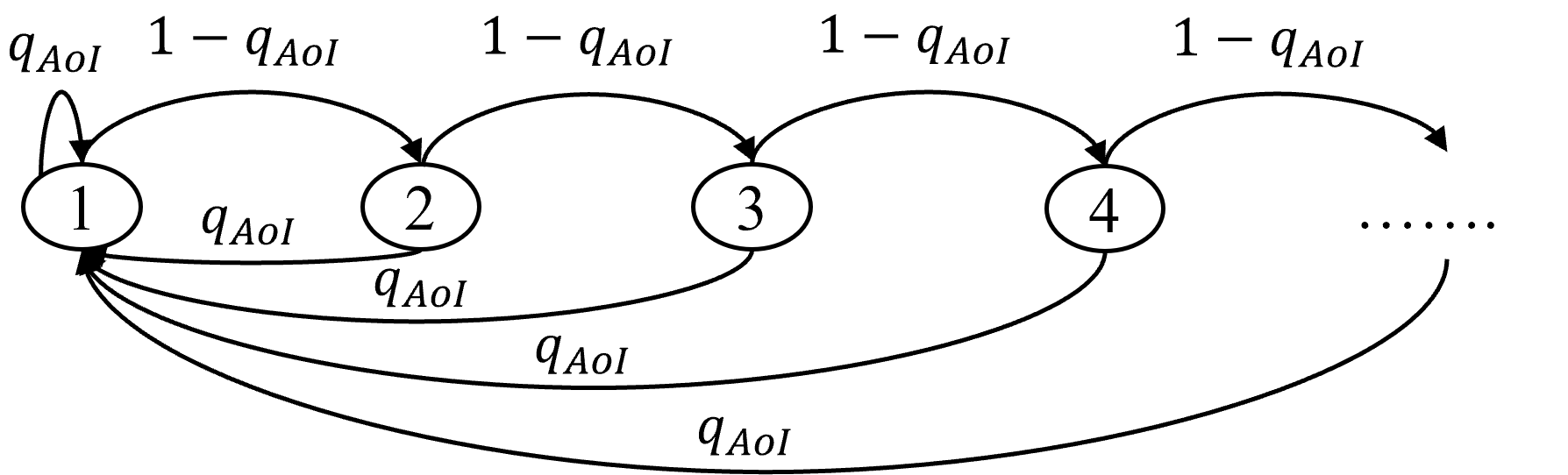}
		\caption{DTMC model of the $\mathrm{T_{AoI}}$ traffic.  \label{DTMC-AoI}}
	\end{figure}
Fig.~\ref{DTMC-AoI} shows the Discrete-Time Markov Chain (DTMC) model of the AoI corresponds to an arbitrary $\mathrm{T_{AoI}}$ node where each state represents the AoI at the central controller. As illustrated in Fig.~\ref{DTMC-AoI}, the AoI-based DTMC transits from any state $n$ to 1 only upon a successful reception of a $\mathrm{T_{AoI}}$ packet, otherwise it transits to state $n+1$. Let $X_t$ represents the value of $\Delta (t)$ at time slot $t$, then the transition probability from state $n$ to state $m$ is $P_{nm}=\mathds{P}(X_{t+1}=m\,\,|X_{t}=n)$, and the transition matrix $\mathbf{P_{AoI}}$ is written as
\begin{equation}
\mathbf{P_{AoI}}=
\begin{bmatrix}
q_{{AoI}} & 1-q_{{AoI}} & 0 & 0 & \dots\\
q_{{AoI}} & 0 & 1- q_{{AoI}} & 0 & \dots\\
\vdots & \vdots & \vdots & \ddots & \ddots
\end{bmatrix}.  
\end{equation}
The row vector $\boldsymbol{\pi^{AoI}}=[\pi_1, \pi_2, ......, \pi_{n-1}, \pi_n, ...]$ represents the steady-state probability vector of the DTMC in Fig.~\ref{DTMC-AoI}, where $\pi_n=\lim_{t \to +\infty}\mathds{P}(X_t)=n$ denotes the probability that the AoI is equal to $n$ at the steady state.
Using the set of equations $\boldsymbol{\pi^{AoI}}\mathbf{P_{AoI}}=\boldsymbol{\pi^{AoI}}$ and $\sum_{n}\pi_n=1$, $\pi_n$ is obtained as 
\begin{equation}
\pi_n=q_{AoI}(1-q_{{AoI}})^{(n-1)}    
\end{equation}
Accordingly, the average AoI $\overline{\Delta}$ is calculated as
\begin{equation}\label{delta_AoI}
\begin{split}
\overline{\Delta}&=\sum_{n=1}^{\infty}\pi_n n=\sum_{n=1}^{\infty} nq_{{AoI}}(1-q_{{AoI}})^{(n-1)}\\
&=\frac{q_{{AoI}}}{1-q_{{AoI}}}\sum_{n=1}^{\infty}n(1-q_{{AoI}})^{n}.
\end{split}
\end{equation}
Since we have $q_{{AoI}}<1$, \eqref{delta_AoI} can be rewritten as   
\begin{equation}
  \overline{\Delta}=\frac{q_{{AoI}}}{1-q_{{AoI}}}\frac{1-q_{{AoI}}}{q_{{AoI}}^2} =\frac{1}{q_{{AoI}}}. 
\end{equation}
However, the average AoI cannot account for extreme AoI events occurring with very low probabilities at IWSNs.
As mentioned in Section~\ref{sec:intro}, the received updates of the monitoring traffic ($\mathrm{T_{AoI}}$) are used for control and actuation actions, which implies certain requirements on tolerated values of AoI.
Here we analyze the AoI violation probability, which is the probability that the AoI exceeds a certain constraint, which can be expressed as
\begin{equation}\label{AoI-violation}
\begin{split}
   &\mathds{P}(\Delta>c)=1-\mathds{P}(\Delta\leq c)=1-\sum_{n=1}^{c}\pi_n\\
   &=\frac{q_{AoI}}{1-q_{AoI}}\sum_{n=1}^{c} (1-q_{AoI})^n=(1-q_{AoI})^c, 
\end{split}
\end{equation}
where $c$ is the target AoI constraint, which is specified according to a considered application scenario.
    \begin{figure}[t!] 
		\centering
		\includegraphics[width= 0.8\linewidth]{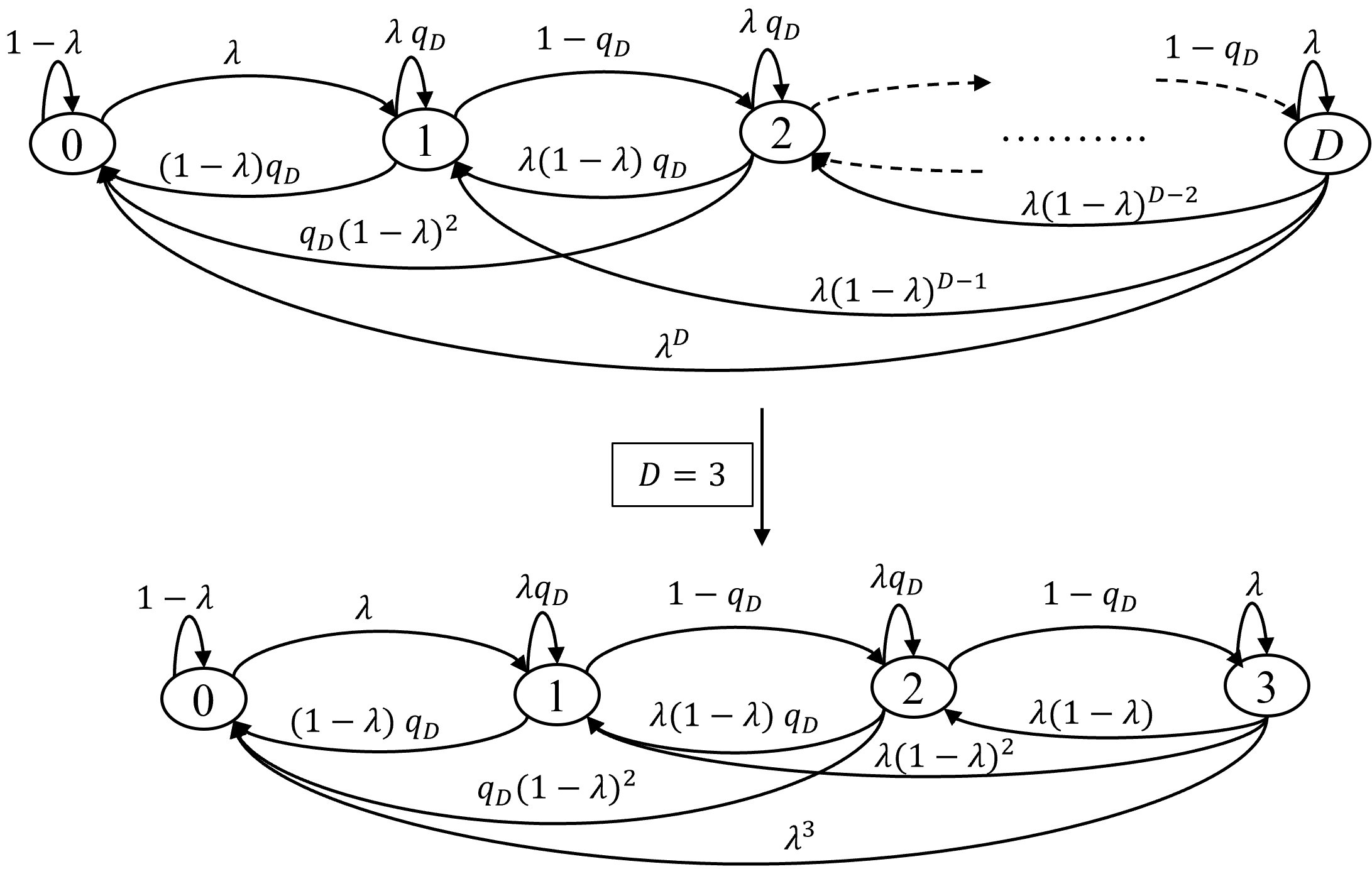}
		\caption{DTMC model of the $\mathrm{T_{D}}$ traffic.  \label{DTMC-deadline}}
	\end{figure}
\setcounter{equation}{12}
	\newcounter{storeeqcounter}
	\newcounter{tempeqcounter}
	\addtocounter{equation}{1}%
	\setcounter{storeeqcounter}%
	{\value{equation}}
	\begin{figure*}[!t]
		\normalsize
		\setcounter{tempeqcounter}{\value{equation}} 
		
		\begin{IEEEeqnarray}{rCl} 
			\setcounter{equation}{\value{storeeqcounter}} 
			\label{deadline-3}
\mathbf{P_{(D=3)}}=
\begin{bmatrix}
(1-\lambda) & \lambda & 0 & 0\\
(1-\lambda)q_{D} & \lambda q_{D} & (1-q_{D}) & 0\\
q_{D}(1-\lambda)^2 & \lambda(1-\lambda)q_{D} & \lambda q_{_D} & (1-q_{D})\\
(1-\lambda)^3 & \lambda(1-\lambda)^{2} & \lambda(1-\lambda)  & \lambda
\end{bmatrix}. 
		\end{IEEEeqnarray}

		\begin{IEEEeqnarray}{rCl}
			\setcounter{equation}{14}\label{deadline-D}
\mathbf{P_D}=
\begin{bmatrix}
(1-\lambda) & \lambda \\
(1-\lambda)q_{D} & \lambda q_{D} & (1-q_{D}) \\
q_{D}(1-\lambda)^2 & \lambda(1-\lambda)q_{D} & \lambda q_{D} & (1-q_{D})\\
\vdots & \vdots & \vdots & \ddots & \ddots \\
q_{D}(1-\lambda)^{D-1} & q_{D}\lambda(1-\lambda)^{D-2} & q_{D}\lambda(1-\lambda)^{D-3} & \dots & \lambda q_{D} & (1-q_{D})\\
(1-\lambda)^D & \lambda(1-\lambda)^{D-1} & \lambda(1-\lambda)^{D-2} & \dots &\lambda(1-\lambda) & \lambda
\end{bmatrix}.  
		\end{IEEEeqnarray}
		\setcounter{equation}{\value{tempeqcounter}} 
		\hrulefill
	\end{figure*}
	\setcounter{equation}{14}
\section{Queue Size and Packet Loss Probability of $\mathrm{T_{D}}$ Traffic} \label{sec:PLP-analysis}
The successful update probability $q_{AoI}$ in \eqref{service-AoI} and \eqref{AoI-violation} is based on the queue status of the $\mathrm{T_{D}}$ user $(\mathds{P}(Q>0))$.
In this section, we use a DTMC model to evaluate the distribution of the queue size of the $\mathrm{T_{D}}$ node.
The DTMC model is shown in Fig.~\ref{DTMC-deadline}, where each state represents the waiting time (in number of time slots) of the Head-of-Line (HoL) $\mathrm{T_{D}}$ packet since its arrival.
The HoL packet is kept waiting in the queue until it is successfully received, i.e., its acknowledgment is received, hence the waiting time accounts for the time slot in which the packet is transmitted.
Since the $\mathrm{T_{D}}$ packet is dropped if its waiting time exceeds the predefined constraint $D$, we have a finite DTMC with $D+1$ states, and the state transitions can be illustrated as follows. 

Assuming that $D=3$ in the example in Fig.~\ref{DTMC-deadline}, there are 4 states for the DTMC.
The state 0 represents empty queue as there is no packet waiting, and remains in the same state as long as there is no arrivals.
The system transits from state 0 to state one when a packet arrives (the HoL packet has the chance to be delivered within one time slot). 
The system remains in state 1 when the HoL packet is successfully transmitted and a new packet arrives ($\lambda q_{D}$). 
A transition from state 1 to state 0 occurs when the HoL packet is successfully transmitted and no packet arrives ($(1-\lambda)q_{D}$). 
The system transits from state 1 to state 2 when the HoL is not successfully transmitted. 
The system remains in state 2 when the HoL packet is transmitted successfully and a new packet arrived in the previous slot ($\lambda q_{D}$). 
A transition from state 2 to state 1 occurs when the HoL is successfully transmitted and one packet arrives in the current slot while no packet arrived in the previous slot ($\lambda(1-\lambda)q_{D}$). 
The system transits from state 2 to state 0 when the HoL packet is successfully transmitted and there were no arrivals within the previous two slots ($q_{D}(1-\lambda)^2$). 
A transition from state 2 to state 3 occurs when the HoL packet is not successfully transmitted, and the packet is now dropped. 
The system remains in state 3 when the HoL packet arrived 3 slots before the current slot. 
The system transits from state 3 to state 0 when no packet arrived within the previous three slots. 
A transition from state 3 to state 1 occurs when a packet arrived in the current slot and no packets arrived in the previous two slots. 
Finally, the system transits from state 3 to state 2 when a packet arrived in the previous slot and no packets arrived in the previous two slots. The transition matrix for the DTMC with $D=3$ can be given as \eqref{deadline-3}.
For an arbitrary deadline constraint $D$, the general transition matrix is expressed as \eqref{deadline-D}.  

The vector $\boldsymbol{\pi^{D}}=[\Grave{\pi_0}, \Grave{\pi_1}, ......, \Grave{\pi_{D-1}}, \Grave{\pi_D}]$ represents the steady-state probability vector of the general DTMC in Fig.~\ref{DTMC-deadline}.
$\boldsymbol{\pi^{D}}$ can be derived from the set of linear equations $\boldsymbol{\pi^{D}}\mathbf{P_D}~=~\boldsymbol{\pi^{D}}$ and $\sum_{n=0}^{D}\Grave{\pi_n}=1$. 
Then, we have $\mathds{P}(Q>0)=1-\Grave{\pi_0}$. 
Moreover, the reliability of the $\mathrm{T_{D}}$ traffic can be represented by $PLP$ with $PLP=\Grave{\pi_D}(1-q_{D})$, which also reflects the deadline violation probability. 
\section{Results and Discussion}
\label{sec:results}
In this section, we evaluate the $\Delta$-$PLP$ trade-off of the considered heterogeneous network based on the presented analysis in Section~\ref{sec:AoI-analysis} and Section~\ref{sec:PLP-analysis}. 
We also validate our analysis by comparing the numerical results with simulation results obtained via discrete-event simulations in MATLAB. 
In the following results, we consider one $\mathrm{T_{D}}$ node located at distance 30~m from the central controller and 5 $\mathrm{T_{AoI}}$ nodes distributed in an isotropic directions around the central controller with equal distance of 40~m.
We set the transmission power for all nodes to 10~dbm and the receiver noise power to -80~dbm.  
    \begin{figure}[t!] 
		\centering
		\includegraphics[width= 0.9\linewidth]{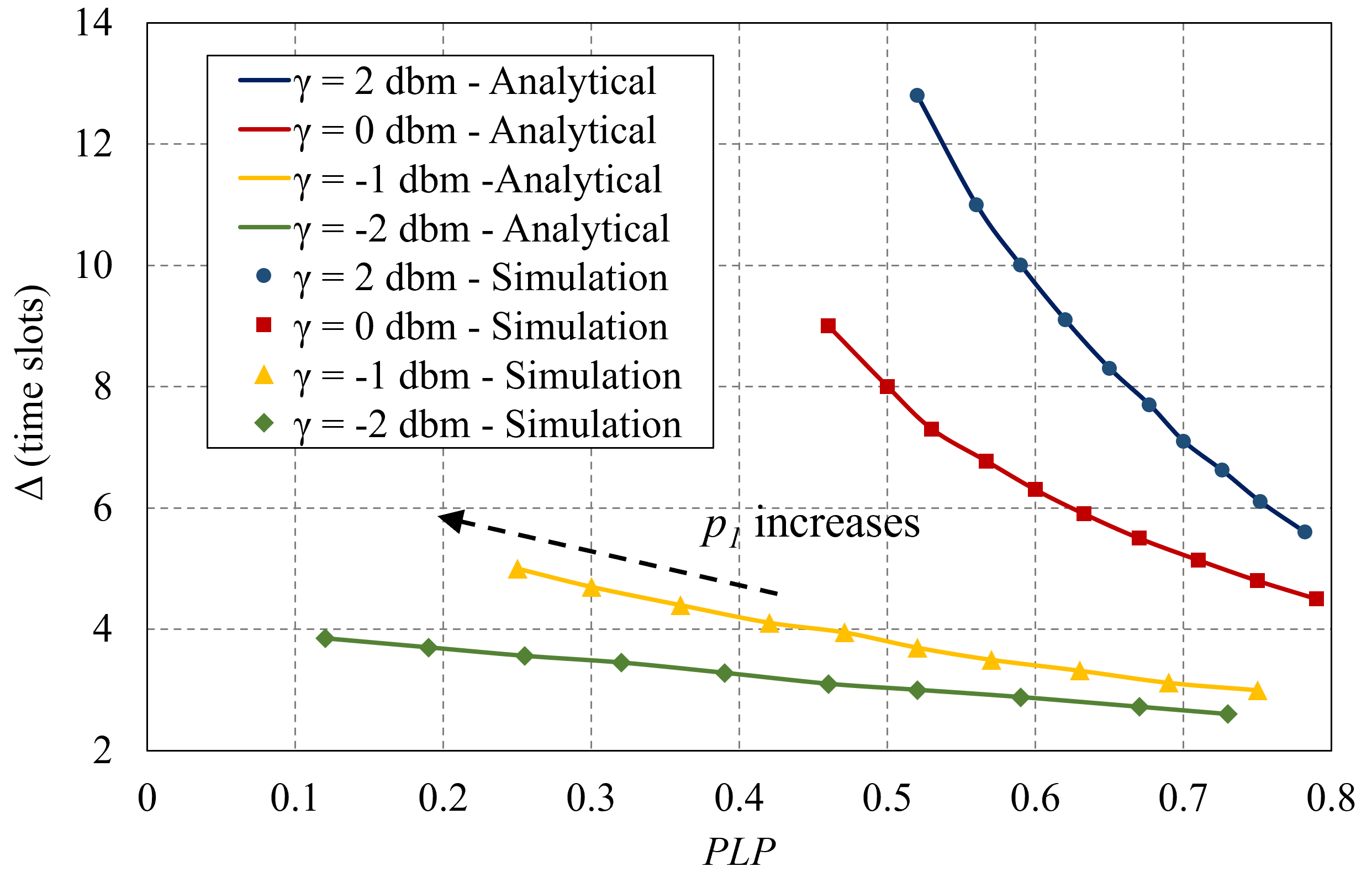}
		\caption{The effect of $p_1$ on $\Delta$-$PLP$ tradeoff under varying $\gamma$ with $0.1\leq p_1\leq1$, $p_2 = 0.6$ and $\lambda = 0.7$. \label{AoI-PLP-p1}}
	\end{figure}

\begin{figure}[t!] 
		\centering
		\includegraphics[width= 0.9\linewidth]{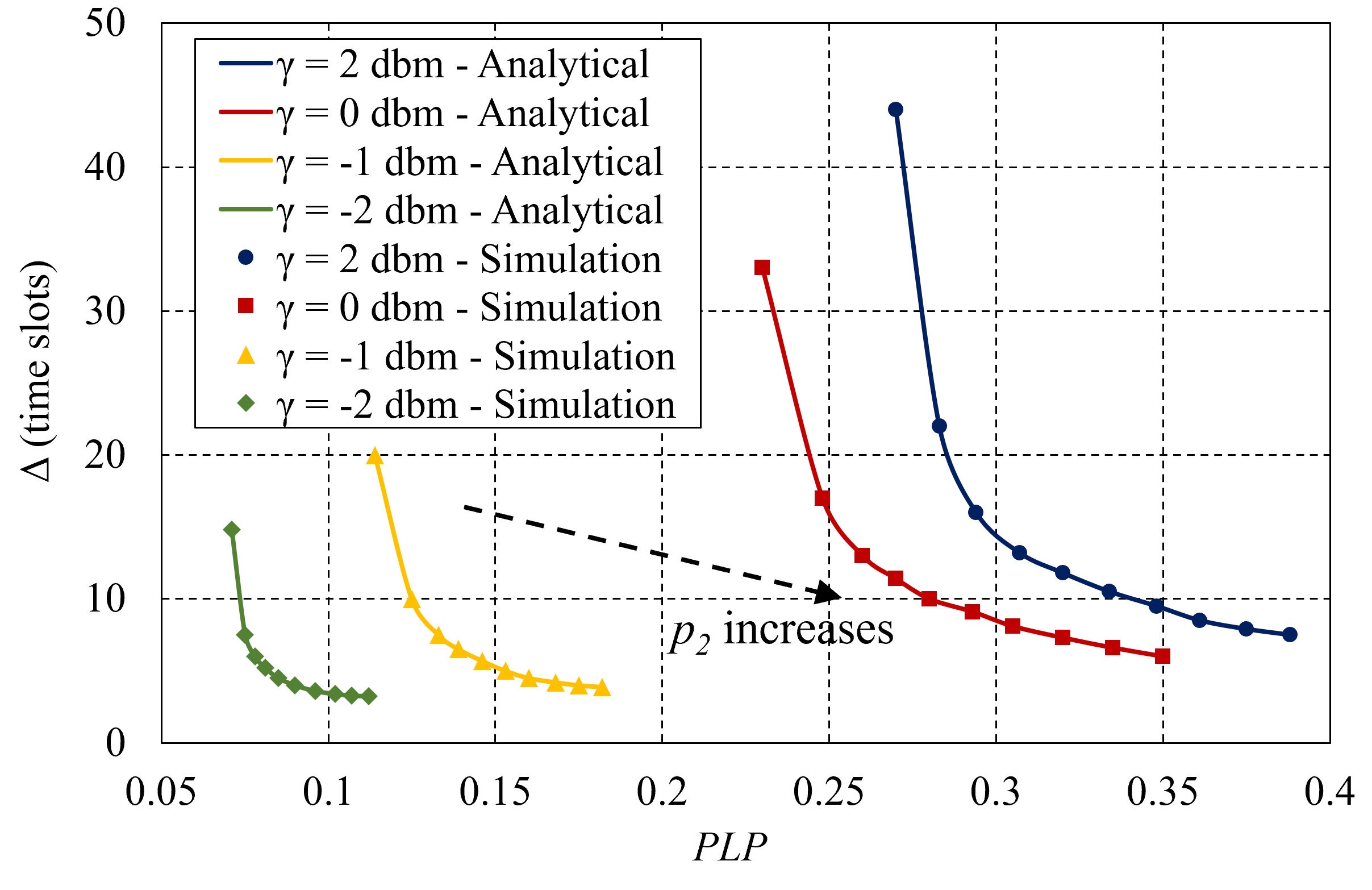}
		\caption{The effect of $p_2$ on the AoI and PLP trade-off under varying $\gamma$ with $p_1 = 0.6$, $0.1\leq p_2\leq1$ and $\lambda = 0.5$.  \label{AoI-PLP-p2}}
	\end{figure}

Fig.~\ref{AoI-PLP-p1} and Fig.~\ref{AoI-PLP-p2} show the evaluation of $\Delta$ and $PLP$ for different values of capture threshold $\gamma$ with varying $p_1$ and $p_2$, respectively. 
From these figures, we can see that the analytical results match well with the simulation results, which validates our analysis in Section~\ref{sec:AoI-analysis} and Section~\ref{sec:PLP-analysis}.
In Fig.~\ref{AoI-PLP-p1} we plot the average AoI ($\Delta$) and $PLP$ by varying $0.1\leq p_1\leq1$ with 0.1 increment. From this figure, we can observe that with high capture capability of the central controller (i.e., $\gamma = -2$ dbm and $\gamma = -1$ dbm), the transmission probability of the $\mathrm{T_{D}}$ node has insignificant effect on the AoI of the $\mathrm{T_{AoI}}$ nodes, while for low capture capability (i.e.,  $\gamma = 0$  dbm and $\gamma = 2$ dbm), the AoI of the $\mathrm{T_{AoI}}$ nodes is highly affected  by higher values of $p_1$. 
For instance, with $\gamma=-2\,$dbm, the AoI increases by 38\% when $p_1$ changes from 0.1 to 0.8, while it increases by 120\% at $\gamma=2\,$dbm.

In Fig.~\ref{AoI-PLP-p2}, we show the $\Delta$-$PLP$ tradeoff by fixing $p_1$ to 0.6 and varying $p_2$ under different values of $\gamma$. We can observe that for low values of $\gamma$, we can decrease $\Delta$ while maintaining low $PLP$ (below 0.2) for the $\mathrm{T_{D}}$ node. On the other side, at high values of $\gamma$, the increase in $p_2$ would decrease $\Delta$ but at the cost of higher $PLP$ of the $\mathrm{T_{D}}$ node. The figure can be used as a reference to select the best transmission strategy to achieve a target pair of $\Delta$ and $PLP$. For instance, with $\gamma=-1\,$dbm, a target AoI below 10 can be achieved at $p_2=0.3$ and $PLP$ = 0.133. Therefore, the transmission strategy in this case would be to allow the $\mathrm{T_{AoI}}$ nodes and the $\mathrm{T_{D}}$ node transmit together. With $\gamma=2\,$dbm, to keep  $\Delta$ below 10, the transmission probability $p_2$ increases to 0.7 while $PLP$ increases to 0.36, meaning that more $\mathrm{T_{D}}$ packets missing the deadline and dropped. In this case, it would be better to adopt a scheduled access strategy (e.g., round-robin) to achieve the target AoI while maintaining low PLP. 
    \begin{figure}[t!] 
		\centering
		\includegraphics[width= 0.9\linewidth]{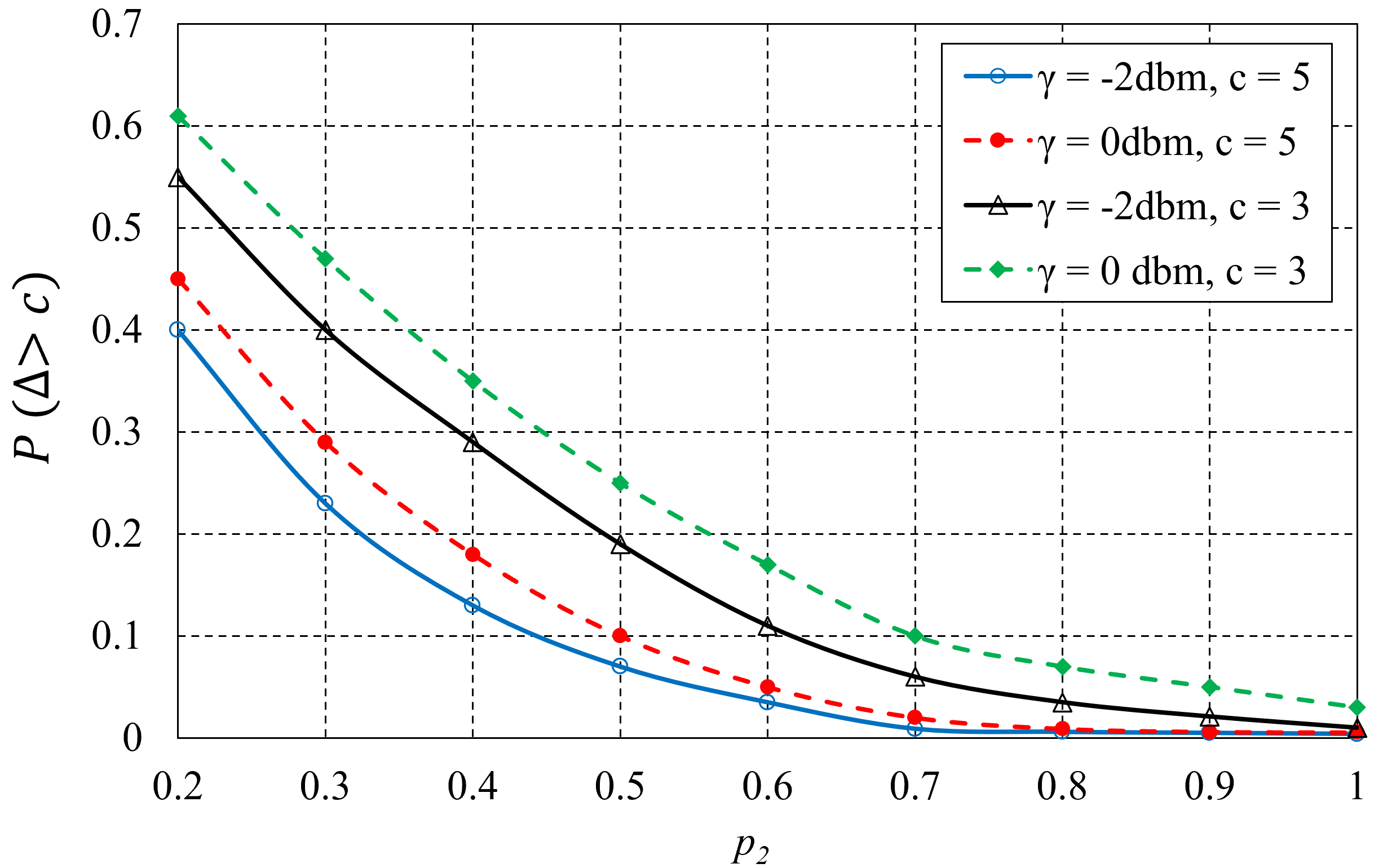}
		\caption{The AoI violation probability under varying $p_2$ with $p_1=0.5$ and $\lambda=0.5$.  \label{AoI-violation}}
	\end{figure}

Fig.~\ref{AoI-violation} shows the AoI violation probability over time obtained from \eqref{AoI-violation} with the target AoI constraints $c=3\,\textrm{and}\,5$ and $\gamma=-2\,\textrm{dbm}\,\textrm{and}\, 0\,\textrm{dbm}$.
As we can see, the AoI violation probability decreases as $p_2$ increases as the $\mathrm{T_{AoI}}$ node attempts to transmit its packet more often. 
Moreover, the AoI violation probability increases significantly for high values of $\gamma$ as it is likely that a $\mathrm{T_{AoI}}$ packet is lost due to the low capture capability. 
The figure gives a more detailed view of the performance when the target is to keep the AoI below a certain constraint. 

\section{Conclusion}
\label{sec:conclusions}
In this paper, we  investigated the AoI-reliability tradeoff in heterogeneous IIoT networks. Specifically, we considered two traffic flows transmitted via unreliable multi-access channel to a central controller, where one flow represents deadline-oriented traffic and the other flow represents AoI-oriented traffic. We derived the average AoI for the AoI-oriented traffic and the PLP for the deadline-oriented traffic using DTMC and validated our analysis through simulations. The obtained results showed that the AoI-PLP tradeoff is mainly influenced by the access probabilities and give insights on how to configure the heterogeneous network to achieve a target AoI-PLP performance. As future work, the analysis in this work can be extended and utilized to propose channel access and /or queue management strategies that improve the AoI performance while maintaining low PLP for more general network setups.   
\section*{Acknowledgement}
This paper has received funding from the European Union’s Horizon 2020 research and innovation programme under grant agreement No. 883315.

	\bibliographystyle{IEEEtran}
\bibliography{mybib}
	
\end{document}